\documentclass
[twocolumn,showpacs,preprintnumbers,floats,amsmath,amssymb]{revtex4-1}
\usepackage{graphicx}
\usepackage{dcolumn}
\usepackage{bm}
\usepackage{amsmath}
\usepackage{amsfonts}
\usepackage{amssymb}
\providecommand{\U}[1]{\protect\rule{.1in}{.1in}}
\begin{document}
\title{Crossing the phantom divide with dissipative normal matter in the Israel-Stewart formalism}

\author{Norman Cruz}
\altaffiliation{norman.cruz@usach.cl}

\affiliation{Departamento de F\'{\i}sica, Universidad de Santiago,
Casilla 307, Santiago, Chile. \\}

\author{Samuel Lepe}
\altaffiliation{samuel.lepe@pucv.cl}

\affiliation{Instituto de F\'{\i}sica, Facultad de Ciencias,
Pontificia Universidad Cat\'olica de Valpara\'\i so, Avenida Brasil
4950, Valpara\'\i so, Chile.\\}

\date{\today}

\begin{abstract}
A phantom solution in the framework of the causal Israel-Stewart
(IS) formalism is discussed. We assume a late time behavior of the
cosmic evolution by considering only one dominant matter fluid with
viscosity. In the model it is assumed a bulk viscosity of the form
$\xi= \xi_{0}\rho^{1/2}$, where $\rho$ is the energy density of the
fluid. We evaluate and discuss the behavior of the thermodynamical
parameters associated to this solution, like the temperature, rate
of entropy, entropy, relaxation time, effective pressure and
effective EoS.  A discussion about the assumption of near
equilibrium of the formalism and the accelerated expansion of the
solution is presented. The solution allows to cross the phantom
divide without evoking an exotic matter fluid and the effective EoS
parameter is always lesser than $-1$ and time independent. A future
singularity (big rip) occurs, but different from the Type I (big
rip) solution classified in S. Nojiri, S. D. Odintsov and S.
Tsujikawa, Phys. Rev. D 71, 063004 (2005), if we consider others
thermodynamics parameters like, for example, the effective pressure
in the presence of viscosity or the relaxation time.
\end{abstract}


\maketitle

\section{Introduction}\label{intro}
A late cosmic stages, phantom phase with an EoS $\omega<-1$~
\cite{Ade2014} is not ruled out by the observational data. A phantom
scheme around today implies future singularities~\cite{Nojiri2005}
and moreover, those singularities could be due to the presence of
(bulk)viscosity~\cite{Cruz2007}. If the EoS of the dark energy is
assumed phantom then there is a violation of the dominant energy
condition (DEC), since $\rho+ p <0$. The energy density grows up to
infinity in a finite time, which leads to a big
rip~\cite{Cadwell},~\cite{Nojiri2004}.

Nevertheless, in the context of matter creation is possible to
explain a phantom behavior without the need of invoking phantom
scalar fields and any modifications in the gravity theory via
dissipative effects~\cite{Nunes}.

The possibility to cross the phantom divide with non ideal fluids
was also derived in the framework of general scalar fields theories.
In~\cite{Vikman} was found that the phantom crossing of the dark
energy described by a general scalar-field Lagrangian is unstable
with respect to the cosmological perturbations. This is the case if
the dominant scalar field is described by the action without
interactions with other energy components throughout kinetic
couplings and higher derivatives. It was also proved that for
general k-essence models the crossing of the phantom divide causes
infinite growth of quantum perturbations on short
scales~\cite{Vikman1}.  Nevertheless, if higher derivatives are add
to the action a single scalar field can cross the phantom divide
without gradient instabilities, singularities or ghosts. This scalar
field corresponds to a velocity potential of an imperfect fluid, and
in an expansion around a perfect fluid it can identified terms which
correct the pressure in the manner of bulk viscosity~\cite{Vikman2}.

The advantage to take a non perfect fluid is that dissipation within
the cosmic fluids allows also a violation of DEC~\cite{Barrow 2} but
the dark fluid do not need to be phantom. In this case we have an
effective pressure given by
\begin{eqnarray}
\label{peff} p_{eff}=p+\Pi ,
\end{eqnarray}
where $p=\omega \rho >0$, being $p$ the barotropic pressure, $\rho $
the energy density and $\Pi <0$ is the viscous pressure. So, we
write $p_{eff}=\left( \omega +\Pi /\rho \right) \rho =\omega
_{eff}\rho $ and $\omega _{eff}$ it could become negative and then
play the role of dark energy~\cite{Velten2013}. For instance, if we
do $\omega =0$ (dust) and by considering for $\Pi $ the model $\Pi
=-3\xi \left( \rho \right) H$ (see later) we have $p_{eff}=\omega
_{eff}\rho $, where $\omega _{eff}=-3\left[ \xi \left( \rho \right)
H/\rho \right] $ and here, $\xi \left( \rho \right) >0$ is the bulk
viscosity coefficient. In goods beads, it is possible to have
phantom evolution driven by viscous dark matter and we do not need
dark energy characterized by $\omega <-1$. And another question is
if the current $\omega $-observational data can be interpreted as
$\omega _{eff}$ or whether we can detect directly parameters
associated to viscosity~\cite{Floerchinger2015}.

The possibility of explain the accelerated expansion of the universe
at late times as an effect of the effective negative pressure due to
bulk viscosity in the cosmic fluids was first considered in
~\cite{Zimdahl}, ~\cite{Balakin}

In the Eckart approach~\cite{Eckart1940}, where the bulk viscosity
introduces dissipation by only redefining the effective pressure of
the cosmic fluid as in Eq.(\ref{peff}), with $\Pi = -3\xi H$, the
possibility of crossing the phantom divide has been found
in~\cite{Brevik}. The magnitude of the viscosity to achive this
crossing using cosmological data was evaluated in~\cite{Brevik1}.
Other investigation in the framework of this approach have explored
big rip singularities for various forms of the EoS parameter and the
bulk viscosity~\cite{Brevik2}, little rip
cosmologies~\cite{Brevik3},~\cite{Gorbunova}; phantom crossing in
modified gravity~\cite{Brevik4},~\cite{Brevik5}; and unified dark
fluid cosmologies~\cite{Li},~\cite{Setare}. At the level of
background evolution a degeneracy between a phantom cosmological
model and $\Lambda$CDM scenarios with a component with bulk
viscosity was studied in~\cite{Velten2013}. The inhomogeneous EOS
for dark energy, where the dependence from Hubble parameter is
included in EOS, has been motivated from the possibility to include
a time-dependent bulk viscosity~\cite{Brevik4}. A study of the
phantom crossing via these EoS was realized and confronted with
astronomical data in~\cite{Nojiri},~\cite{Bamba}

It is a well known results that the Eckart approach has drawbacks
related to causality and stability. Nevertheless, it is a reasonable
assumption, for example, at early times, if we are thinking in
"short distances" between interacting cosmic components, i.e.,
almost instantaneous propagation between them. So, a null relaxation
time is a good setting. At late times this non-causal approach does
not work, obviously, and this is the main reason for considering
causal approaches for bulk viscosity (non-null relaxation time).
Previous investigations consider the Eckart approach as a first
simple approach to study viscous cosmologies. For example, the
thermal fluctuations in the very early stage of viscous cosmology
and the spectral index and non-gaussianity was studied in
~\cite{Wei-Jia}. Also, the structure formation in a viscous CDM
universe was considered in~\cite{Velten2014}. The Eckart and causal
formalism was used to face the problems at non-linear level of the
$\Lambda CDM$ model in~\cite{Barbosa}, where authors proposed a
viscous model with cosmological constant, which not present, in
principle, drawbacks, but reject viscous cosmologies as viable
framework for unified models of the dark sector.

The investigation on the nature of phantom behavior from dissipative
process would not be complete understood without taking into account
a more physical approach like the full Israel-Stewart (IS) causal
thermodynamics. Since in this framework there is a great difficulty
to obtain solutions to the main equations, only some partial results
have been found. In the special case where the bulk viscosity
coefficient takes the form $\xi (\rho) \sim \rho^{1/2}$, a big rip
singularity solution was obtained in this formalism for a late time
FRW flat universe filled with only one barotropic fluid with bulk
viscosity~\cite{Cataldo2005}. Nevertheless, this solution was
obtained in the linear IS theory which relies on the assumption of
small deviations from thermodynamics equilibrium,i.e., $|\Pi| <p$.
This assumption is not hold in the case of accelerated expansion, as
the observed at late times of the cosmic evolution or during
inflation.

Our aim in this work is to study the thermodynamical properties of
the phantom solution that is also obtained in the IS formalism when
a more consistent expression for the relaxation time is taken into
account, derived from the speed of bulk perturbations (see~
\cite{Maartens1996}) and showed in Eq.(\ref{relaxationtime}). In the
big rip solution found in~\cite{Cataldo2005} the relaxation time
$\tau$ was defined as $\xi/\rho $, where $\xi$ is the bulk viscosity
coefficient and $\rho$ is the energy density of the matter
component. The main result was that the EoS of the barotropic fluid
with bulk viscosity must be of phantom type. In other words there is
no crossing of the phantom divide due to the viscosity in the full
causal formalism, when the above expression for $\tau$ is assumed.
Nevertheless, in the big rip solution present here the EoS parameter
of the fluid is in the range $0<\omega<1/2$ and the effective EoS
due to the presence of viscosity correspond to a phantom matter. So
a crossing of the phantom divide due to the viscosity is allowed.
From a theoretical point of view this results seems to show that
\textit{the crossing of the phantom divide were somehow related to
the need to maintain causality}~\cite{Scherrer}. A similar result
was found using the Lichnerowicz approach to
viscosity~\cite{Disconzi}. So, the main motivation to explore
phantom solutions using a causal approach is to investigate the
physical viability  to cross the phantom divide without invoking
phantom fields.

We will explore further this phantom solution evaluating their
entropy generation, temperature of the fluid and viscous bulk
pressure as a function of the cosmic time. We will also discuss the
possibility to extend the classification of singularities given
in~\cite{Nojiri2005} since the singularities obtained with the
inclusion of dissipation must be characterized by also the behavior
of other thermodynamic parameters, like the effective pressure and
the relaxation time.

The organization of the paper is as follows: In Section II we
present a brief revision of the IS formalism and we show the phantom
solution found. In Section III we discuss the near equilibrium
conditions in the case of accelerated expansion. We also explore the
behavior on time of the thermodynamics parameters like the entropy
generation, the viscous pressure, the temperature, the entropy and
the relaxation time. In Section IV we propose to extend the
classification of singularities analize some aspects of the
thermodynamical equilibrium. Section V is devoted to conclusions.
$8\pi G=c=1$ units will be used.

\section{Israel-Stewart formalism} \label{dos}

In what follows we assume only one fluid as the main component of
the universe, which experiment dissipative process during cosmic
evolution. This fluid obey a barotropic EoS, $p=\omega \rho $, where
$p$ is the barotropic pressure and $0\leq \omega <1$. For a flat
FLRW universe, the equation of constraint is
\begin{eqnarray}
\rho =3H^{2}. \label{eq:eq0}
\end{eqnarray}
In the IS framework the transport equation for the viscous pressure
$\Pi $ is given by \cite{Israel1979}
\begin{eqnarray}
\tau \dot{\Pi}+\left( 1+\frac{1}{2}\tau \Delta \right) \Pi
=-3\xi(\rho) H, \label{eq:eq1}
\end{eqnarray}
where "dot" accounts for the derivative with respect to the cosmic
time. $\tau$ is the relaxation time, $\xi (\rho)$ is the bulk
viscosity coefficient which depends on the energy density $\rho$,
$H$ is the Hubble parameter, and $\Delta$ is defined by
\begin{equation}
\Delta \equiv 3H+\frac{\dot{\tau}}{\tau }-\frac{\dot{\xi}}{\xi
}-\frac{\dot{T}}{T}, \label{eq:eq2}
\end{equation}
where $T$ is the barotropic temperature, which takes the form
$T=\beta \rho ^{\omega /\left(\omega +1\right)}$ (Gibbs
integrability condition when $p=\omega \rho $) with $\beta $ being a
positive parameter. We also have that~\cite{Maartens1996}
\begin{equation}
\frac{\xi}{\left(\rho +p\right)\tau} =
c_{b}^{2},\label{relaxationtime}
\end{equation}
where $c_{b}$ is the speed of bulk viscous perturbations
(non-adiabatic contribution to the speed of sound in a dissipative
fluid without heat flux or shear viscosity), $c_{b}^{2}=\epsilon
\left(1-\omega \right)$ and $0<\epsilon \leq 1$ ($\epsilon
=1\Leftrightarrow $ H-theorem, entropy production is non-negative),
$\xi =\xi _{0}\rho ^{s}$ being $\xi _{0}$ a positive constant, if
the second law of thermodynamics is respected~\cite{Weinberg1971}
and can be estimated, for example in the Eckart formalism, from the
observational data~ \cite{Avelino2013}. $s$ is an arbitrary
parameter. So, the relaxation time results to be (and $\epsilon =1$
from now on)
\begin{eqnarray}
\tau =\frac{1}{1-\omega ^{2}}\frac{\xi }{\rho }=\frac{\xi
_{0}}{1-\omega ^{2} }\rho ^{s-1},  \label{eq:eq3}
\end{eqnarray}
and, according to (\ref{eq:eq2})

\begin{eqnarray}
\Delta =\frac{3H}{\delta \left( \omega \right) }\left( \delta \left(
\omega \right) -\frac{\dot{H}}{H^{2}}\right) , \label{eq:eq4}
\end{eqnarray}
where we have defined the $\delta \left( \omega \right) $ parameter

\begin{eqnarray}
\delta \left( \omega \right) \equiv \frac{3}{4}\left( \frac{1+\omega
}{1/2+\omega } \right). \label{eq:eq5}
\end{eqnarray}
So, for $0\leq \omega <1$, $\delta \left( \omega \right) >0$.  Using
Eq.(\ref{eq:eq3}) and Eq.(\ref{eq:eq0}) we can write

\begin{eqnarray}
\tau H=\frac{3^{s-1}\xi _{0}}{1-\omega ^{2}}H^{2\left( s-1/2\right)
}, \label{eq:eq7}
\end{eqnarray}
and we see that $\tau \longrightarrow 0$ when $H$ increase for
$s<1$, in particular if $s=1/2$ and if $s<0$. This idea appears
reasonable if $\dot{\Pi }$ does not increase faster that $\tau $
going to zero and maintaining finite $\Delta $ (for instance, if
$\omega _{eff}\sim -1$ and $0\leq \omega <1$). Negative potencies
appears to be consistent with the observational data in the
Eckart's framework, in particular, $s\leq -1/2$~\cite{Velten2011}.

Finally, by using for $p$, $T$ and $c_{b}^{2}$ the expressions given
before, the equation (\ref{eq:eq1}) can be written in the form

\begin{eqnarray}
\tau _{\ast }\dot{\Pi}+\Pi =-3\ \xi _{\ast }\left( \rho \right)
H\left[ 1+ \frac{1}{1-\omega ^{2}}\left( \frac{\Pi }{\rho}\right)
^{2}\right], \label{eq:eq11}
\end{eqnarray}
where the effective relaxation time $\tau _{\ast }$ and the
effective bulk viscosity $\xi _{\ast }$\ are, respectively,

\begin{eqnarray}
\tau _{\ast }=\frac{\tau }{1+3\left( 1+\omega \right) \tau
H},\label{eq:eq111}
\end{eqnarray}
and
\begin{eqnarray}
\xi _{\ast }=\frac{\xi }{1+3\left( 1+\omega \right) \tau H }=\left(
1-\omega ^{2}\right) \rho \tau _{\ast }. \label{eq:eq12}
\end{eqnarray}

Since we are interested in the possibility of phantom solution in
the framework of dissipative process, we shall also consider the
truncated equation version of Eq. (\ref{eq:eq11}), where the near
equilibrium condition

\begin{eqnarray}
\left\vert \frac{\Pi }{\rho}\right\vert <<1,  \label{eq:eq13}
\end{eqnarray}
allows to neglect the second term in square brackets in Eq.
(\ref{eq:eq11}) and then it reduces to

\begin{eqnarray}
\tau _{\ast }\dot{\Pi}+\Pi =-3\text{\ }\xi _{\ast }\left( \rho
\right) H. \label{eq:eq14}
\end{eqnarray}

The Eckart formalism (non causal formalism)~\cite{Eckart1940} comes
after to set in (\ref{eq:eq14}) $\tau _{\ast }=0$ ($\tau =0$). We
note also that if $\tau H<<1$ then $\tau _{\ast }\approx \tau $,
$\xi _{\ast }\approx \xi $ and (\ref{eq:eq14}) is reduced to

\begin{eqnarray}
\tau \dot{\Pi}+\Pi =-3\text{\ }\xi \left( \rho \right) H.
\label{eq:eq15}
\end{eqnarray}

\subsection{Phantom solution} \label{tres}

In order to find a phantom solution for a universe filled with one
dominant fluid with positive pressure and viscosity we construct a
differential equation for the Hubble parameter. By using the
conservation equation

\begin{eqnarray}
\dot{\rho}+3H\left[ \left( 1+\omega \right) \rho +\Pi \right] =0,
\label{eq:eq18}
\end{eqnarray}
the Eq.(\ref{eq:eq0}) and the relation $\xi \left( \rho \right)
=\xi _{0}\rho ^{s}$ and Eq.(\ref{eq:eq1}), we can obtain the
following differential equation
\begin{widetext}
\begin{eqnarray}
\left[ \frac{2}{3\left( 1-\omega ^{2}\right) }\left( \frac{3\left(
1+\omega \right)\dot{H}}{H^{2}}+\frac{\ddot{H}}{H^{3}}\right)
-3\right] H^{2\left( s-1/2\right) }+
\frac{1}{3^{s}\xi _{0}}\left[
1+\frac{3^{s-1}\xi _{0}\Delta H^{2\left( s-1\right)}}{2\left(
1-\omega ^{2}\right) } \right] \left[ 3\left( 1+\omega \right)
+\frac{2\dot{H}}{H^{2}}\right] = 0. \label{eq:eq19}
\end{eqnarray}
\end{widetext}
In this framework a big rip solution was found for $s=1/2$
in~\cite{Cataldo2005}, using the following Ansatz
\begin{eqnarray}
H\left( t\right) =A\left( t_{s}-t\right) ^{-1}. \label{Ansatz}
\end{eqnarray}
It is easy to verify that for $s=1/2$ a quadratic equation for $
A=const.$ is obtained when Eq.(\ref{Ansatz}) is introduced in
Eq.(\ref{eq:eq19}). The solutions of this equation are detailed in
the Appendix. On the other hand, and as far as we know, for $s\neq
1/2$ (or $s\leq -1/2 $) there is not a phantom solution of this type
in the IS formalism.

Before to discuss the properties of the solution obtained we will
inspect now the possibility of finding a phantom scheme in the
truncated IS formalism, represented through Eq.(\ref{eq:eq14}). By
doing this, we follow the procedure done before. For $s=1/2$, we
obtain the following equation for the Hubble parameter
$\ddot{H}+\alpha H\dot{H}+\beta H^{3}=0$, where $\alpha $ and $
\beta $ are both constants. By replacing the solution $H\left(
t\right) =B\left( t_{s}-t\right) ^{-1}$, we have for $B$ an
algebraic equation for which there is not a positive root and then
there is not a phantom solution. Doing the same thing with the
truncated version of IS formalism represent by (\ref{eq:eq15}), we
find a phantom solution if $\sqrt{3}\xi _{0}> 1+\omega $. Finally,
by using the Eckart scheme we find a phantom solution if
$\sqrt{3}\xi _{0}< 1+\omega  $. So, the constraint over $\xi _{0}$
decides whether there is or not a phantom solution in the considered
formalism.

The above results indicates that phantom solution is obtained in the
non causal framework of Eckart and in the causal formalism of
Israel-Stewart. Nevertheless, being both approaches physically
different it is important to investigate further all the aspects
involved in the causal framework. It is significant that the
truncated version of IS formalism, which assume a near equilibrium
process, do not admits a phantom solution like the Ansatz mentioned
above.
\newpage
\section{Thermodynamical properties of the phantom solution} \label{cuatro}

As we mentioned in the previous section, in the truncated version of
the IS formalism represented by Eq.(\ref{eq:eq14}) there is no
phantom solution of the type $H\left( t\right) =A\left(
t_{s}-t\right) ^{-1}$. Since in this approach the basic equation is
constructed demanding the near equilibrium condition, it is expected
that a phantom solution may have problems with the requirement of
the condition given in the inequality (\ref{eq:eq13}). This point
has been already discussed by Maartens~\cite{Maartens} in the
context of dissipative inflation where the universe is filled by
only one 'ordinary' fluid with positive equilibrium pressure and the
negative effective pressure ($p_{eff} = p+ \Pi$) is due to a viscous
stress. For late times behavior the observed acceleration of the
expansion is a condition that we expect to obtain from this negative
effective pressure, so the condition $\ddot{a}>0$ leads to
\begin{eqnarray}
-\Pi > p +\frac{\rho}{3}. \label{Pieqa}
\end{eqnarray}

So the inequality (\ref{Pieqa}) implies that the viscous stress is
greater than  the equilibrium pressure. The causal approach assume a
near equilibrium regime but in order to obtain accelerated expansion
the fluid has to be far from equilibrium. To overcome this situation
a nonlinear generalization of the causal linear thermodynamics of
bulk viscosity has been implemented in~\cite{MaartensMendez}. We do
not consider here this generalization but we shall show that in the
case of the de Sitter solution there is no difference with the
results obtained in~\cite{MaartensMendez}. In order to explore the
thermodynamics behavior of our phantom solution obtained solving the
Eq.(\ref{eq:eq19}) by means of the Ansatz (\ref{Ansatz}), we will
evaluate the viscous pressure, $\Pi (t)$, and the entropy $S(t)$.

The big rip solution obtained for $s=1/2$ is singular in the sense
that only for this value of the $s$ parameter, Eq. (\ref{eq:eq19})
do not admit a de Sitter solution. It is straightforward to verify
this since that for a de Sitter solution, $H=const.=H_{0}$, Eq.(
\ref{eq:eq19}) becomes in a simple algebraic equation for $H_{0}$,
whose solution is given by
\begin{eqnarray}
H_{0}=\left\{ \frac{1}{3^{s}\xi _{0}}\left( \frac{1-\omega
^{2}}{1/2-\omega } \right) \right\} ^{1/2\left( s-1/2\right) }.
\label{eq:eq21}
\end{eqnarray}
So, the above equation represents a solution of the Hubble parameter
if $s\neq 1/2$ and $0<\omega <1/2$. We can explore first how behaves
in this case $\Pi (t)$ and $S(t)$. We do not have a future-time
singularity and the time derivatives of the Hubble parameter are
zero, so the thermodynamics parameters are more easily evaluated.

Let us begin evaluating $\Pi (t)$. Using Eq.(\ref{eq:eq0}) in the
continuity equation (\ref{eq:eq18}) we can write the following
expression for $\Pi (t)$
\begin{equation}
\Pi =-2\dot{H}-3\left( 1+\omega \right) H^{2}, \label{PIenfuncion
deH}
\end{equation}
so introducing in Eq. (\ref{PIenfuncion deH}) the solution $H=H_{0}$
we obtain
\begin{equation}
\Pi _{0}=-3\left( 1+\omega \right) H_{0}^{2}=const.,
\end{equation}
We examine now the entropy generation and the entropy as a function
of the cosmic time. The entropy generation can be evaluated from the
expression
\begin{equation}
nT\frac{dS}{dt}=-3H\Pi , \label{entropygeneration}
\end{equation}%
where $n$ is the number density of particles, which satisfy the
continuity equation
\begin{equation}
\dot{n}+3Hn=0, \label{continuityn}
\end{equation}
whose solution in terms of the scale factor $a(t)$ is
\begin{equation}
n\left( a\right) =n_{0}\left( a/a_{0}\right) ^{-3}. \label{ndea}
\end{equation}
So using Eqs.(\ref{entropygeneration}) and (\ref{ndea}) and the
expression for the temperature given by $T_{0}=\beta \rho
_{0}^{\omega /\left( \omega +1\right) }=\beta \left(
3H_{0}^{2}\right) ^{\omega /\left( \omega +1\right) }$, we obtain
\begin{equation}
n\left( t\right) =n_{0}\exp \left( -3H_{0}t\right) ,
\end{equation}
and then the entropy as a function of time takes the form
\begin{equation}
S\left( t\right) =const.+3\left( 1+\omega \right)
\frac{H_{0}^{2}}{n_{0}T_{0} }\exp \left( 3H_{0}t\right) .
\end{equation}
Summarizing, de Sitter solution obtained from a causal dissipative
approach can be obtained with a constant viscous pressure with an
exponentially increase of entropy. It is interesting to note that
the relaxation time $\tau $ is also constant. According to Eq.
(\ref{eq:eq3})
\begin{equation}
\tau _{0}=\frac{\xi _{0}}{1-\omega ^{2}}\rho _{0}^{s-1}=
\frac{\xi_{0}}{ 1-\omega ^{2}}\left( 3H_{0}^{2}\right)
^{s-1}=const.,
\end{equation}

The above results has no differences with the obtained
in~\cite{MaartensMendez}, in the framework of a nonlinear
generalization of the causal linear thermodynamics of bulk
viscosity. As it is expect on simple argumentes, the effective EoS
defined by
\begin{equation}
\omega_{eff} = \frac{p(t)+\Pi(t)}{\rho(t)}, \label{omegaefectivo}
\end{equation}
is always equal to $-1$ and independent of EoS parameter $\omega$.

Let us evaluate now the corresponding parameters of our big rip
solution $H\left( t\right) =A\left( t_{s}-t\right) ^{-1}$, which
presents a singularity in its parameters in a finite time. A simple
integration gives us the scale factor as a function of time
\begin{equation}
a/a_{0}=\left( t_{s}-t_{0}\right) ^{A}/\left( t_{s}-t\right) ^{A},
\end{equation}
so the number density of particles yields
\begin{equation}
n\left( t\right) =n_{0}\left( \frac{t_{s}-t}{t_{s}-t_{0}}\right)
^{3A}.
\end{equation}
Of course, at the time $t=t_{s}$ the size of the universe becomes
infinite and the number density of particles goes to zero. The
temperature is given by
\begin{equation}
T\left( t\right) =\beta \left( 3A^{2}\right) ^{\omega /\left( \omega
+1\right) }\left( t_{s}-t\right) ^{-2\omega /\left( \omega +1\right)
}, \label{temperatureenfunciondet}
\end{equation}
and the viscous pressure can be obtained introducing our Ansatz in
Eq.(\ref{PIenfuncion deH}) which yields

\begin{equation} \Pi \left( t\right) =-\left[ 2+3\left( 1+\omega
\right) A\right] A\left( t_{s}-t\right) ^{-2},
\label{PIenfunciondet}
\end{equation}
so the viscous pressure becomes infinity at the singularity
increasing the temperature of the fluid to infinity, as it can be
seen from Eq. (\ref{temperatureenfunciondet}) since the power
$-2\omega /\left( \omega +1\right)$ is always negative for $\omega
>0$.

Let us make some comment about the inequality (\ref{Pieqa}) which is
the condition to have an accelerated expansion. The energy density
of the phantom fluid takes the form
\begin{equation}
\rho (t) = 3A^{2}(t_{s}-t)^{-2}, \label{rhodet}
\end{equation}
so the pressure is $p(t)=\omega \rho (t)$. Then introducing
Eq.(\ref{PIenfunciondet}) and Eq.(\ref{rhodet}) together with the
expression for $p(t)$ in the inequality (\ref{Pieqa}) it is
straightforward to see that the only condition that this inequality
imposes is of $A>0$, which is a condition of our solution.

The increasing rate of entropy can be evaluated from Eq.(
\ref{entropygeneration}) and we obtain the following expression
\begin{equation}
\frac{dS}{dt}=C\left( t_{s}-t\right) ^{\eta },
\label{entropygenerationdet}
\end{equation}
where

\begin{equation}
C =\frac{3A^{2}\left[ 2+3\left( 1+\omega \right) A\right]
}{n_{0}\beta \left( 3A^{2}\right) ^{\omega /\left( \omega +1\right)
}}\left( t_{s}-t_{0}\right) ^{3A}>0,
\end{equation}
and
\begin{equation}
\eta  =\frac{2\omega }{\omega +1}-3(1+A).
\end{equation}

Since the natural tendency of systems to evolve toward
thermodynamical equilibrium is characterized by two properties of
its entropy function: $dS/dt>0$ and $d^{2}S/dt^{2}<0$ ($S$
\textit{is convex}), we also evaluate $d^{2}S/dt^{2}$ in order to
get new possible constraints on the parameters of the model.
Deriving once Eq.(\ref{entropygenerationdet}) we obtain that
\begin{equation}
\frac{d^{2}S}{dt^{2}}=C\eta \left( t_{s}-t\right) ^{\eta -1}
\end{equation}

Then for this phantom solution  $dS/dt>0$ and $d^{2}S/dt^{2}<0$ is
satisfied if $\eta <0$.

Integration of Eq.(\ref{entropygenerationdet}) yields the entropy
as a function of the cosmic time
\begin{equation}
S\left( t\right) =- const.\left (\frac{C}{\eta +1}\right )\left(
t_{s}-t\right) ^{\eta +1},
\end{equation}
so $S\left( t\right) >0$ if  $\eta +1<0$. Then, we have two
conditions for $\eta $ that must be satisfied: $\eta <0$ and $\eta
+1<0$, which reduce to the condition $\eta <-1$, i.e.,$
\frac{2\omega }{\omega +1}-3(1+A)<-1$, which leads to the inequality
\begin{equation}
A>\frac{1}{3}\left( 1+ \frac{2\omega }{\omega +1}\right) -1,
\label{Aineq}
\end{equation}
and since $A>0$ then (\ref{Aineq}) is always verified for the
values $0<\omega < 1/2$. So, our solution verifies naturally the
thermodynamics requirements:  $S>0$, $dS/dt>0$ and
$d^{2}S/dt^{2}<0$.

We evaluate the relaxation time introducing Eq.(\ref{Ansatz}) in
Eq.(\ref{eq:eq3}) for $s=1/2$, which yields
\begin{equation}
\tau \left( t\right) =\frac{\xi _{0}/\sqrt{3}}{\left( 1-\omega
^{2}\right) A} \left( t_{s}-t\right). \label{taudet}
\end{equation}
At the singularity the relaxation time goes to zero. A obvious
question is if the effective EoS due to the viscosity in a fluid of
positive pressure correspond to a phantom matter.  Evaluating the
effective EoS defined in Eq.(\ref{omegaefectivo}) we obtain
\begin{equation}
\omega_{eff} = -1 -\frac{2}{3A}, \label{omegaefectivo1}
\end{equation}
so the effective EoS do not depends on time and is always phantom.

\section{The singularity of the phantom solution} \label{cinco}

The properties of future singularities when the universe is
dominated by a phantom fluid with an EoS of the form
\begin{equation}
p=-\rho -f(\rho),\label{EoSphantom}
\end{equation}
was investigated in~\cite{Nojiri2005}, where a classification in
four types was done. The function $f(\rho)$ can be an arbitrary
function but the dominant energy condition (DEC) is always violated.
It has been pointed out that this EoS may be equivalent to bulk
viscosity~\cite{Barrow 2}, nevertheless the physical behavior of
this EoS in the framework of a perfect phantom fluid and a non
phantom fluid with a bulk viscosity in the IS formalism are quite
different. In this sense, our solution presents a Type I singularity
(Big Rip) if we follow the classification given
in~\cite{Nojiri2005}, where for $t\rightarrow t_{s}$, $a\rightarrow
\infty$, $\rho\rightarrow \infty$, and $|p|\rightarrow \infty$. In
our case we have other parameter like the viscous pressure which
also can be considered in the characterization of the singularity.
For our solution this parameter also diverges at the singularity.
But more important is the effective pressure defined in
Eq.(\ref{peff}) since drives the effective EoS of the viscous fluid.
Due to non perfect fluids include the effective pressure as a new
parameter to be evaluated at the singularity, we propose to extend
the classification proposed in~\cite{Nojiri2005} in order to taken
into account this new behavior for solutions that present
singularities in the presence of viscosity. Specifically, we propose
to define naturally the Type $I^{\ast}$ (Viscous Big Rip) future
singularity for non perfect fluids. This singularity can be
characterized in the following way: for $t\rightarrow t_{s}$
\begin{equation}
a\rightarrow \infty, \rho\rightarrow \infty, |p|\rightarrow \infty,
|p_{eff}|\rightarrow \infty,\label{parametersatinifinity}
\end{equation}
and the higher temporal derivatives of $H$ also diverge. Note that
$p_{eff}$ in terms of $H$ and $\dot{H}$ is given by the expression
\begin{equation}
p_{eff}=-2\dot{H}-3H^{2}, \label{peffenfuncion deH}
\end{equation}
so the divergence of $H$ at the singularity implies directly the
divergence of $p_{eff}$ in both thermodynamical approaches of Eckart
and IS.

The divergence of the Hubble rate (Eq.\ref{Ansatz}) also implies the
divergences of all curvatures. An special feature of our solution is
the constancy of the effective EoS given in
Eq.(\ref{omegaefectivo1}), which is always phantom. In general, in
the framework of perfect fluids obeying the EoS given by
Eq.(\ref{EoSphantom}) the effective EoS is constant only in some
regions like $t\ll t_{s}$ or $t\sim t_{s}$ (see for example the
solution $H(t)= n (1/t +1/(t_{s}-t))$, where n is a positive
constant, in~\cite{Nojiri2005}). Additionally, note that despite the
divergences in the parameters showed in
Eq.(\ref{parametersatinifinity}) the relaxation time given by
Eq.(\ref{taudet}) goes to zero at the singularity. Since this
parameter also characterize the thermodynamic behavior of our
solution, could be included in the characterization of the
singularities.

Other cosmological solutions of universes filled with dissipative
fluids, which presents future singularities were found
in~\cite{Cruz2007}. In this case the dark energy component was
assumed to be a generalized dissipative Chaphygin. When the
dissipative effects were evaluated in the framework of the
non-causal Eckart theory, the barotropic pressure, $p$, satisfies
$\left\vert p\left( t\longrightarrow t_{s}\right) \right\vert
\longrightarrow 0$, but the effective pressure goes to infinity at
the singularity. For the case where the dissipative effects were
analyzed in the truncated version of the IS formalism, we obtained
solutions with the following behavior for $t\rightarrow t_{s}$:
\begin{equation}
a\rightarrow \infty, \rho\rightarrow \infty, |p|\rightarrow \infty,
|p|\rightarrow constant,\label{parametersatinifinityChapligyn}
\end{equation}
where $|p_{eff}|\rightarrow constant$ also, since the bulk pressure
$\Pi$ always satisfy the condition $|\Pi| \ll |p|$. In one of the
solutions the constant is zero.

So, these solutions are not included in the classification above
suggested and it is an indication that the classification for
singularities occurring with dissipative fluids should be extended.

\section{Final remarks} \label{cinco}

We have discussed in the framework of IS causal formalism the
thermodynamics properties of a big rip solution of the type
$H=A(t_{s}-t)^{-1}$. This solution was found for a flat FRW universe
filled with a barotropic fluid with an EoS parameter $\omega
>0$. Our solution implies a cosmological scenario where a barotropic fluid with $0<\omega<1/2$
behaves like a phantom fluid with constant EoS $\omega<-1$, driving
by the viscosity. So, this solution allows to cross the phantom
divide with normal matter in the full causal formalism of IS. In the
previous phantom solution found in~\cite{Cataldo2005} there is no
phantom crossing, since the corresponding dissipative fluid must
have a phantom EoS from the beginning. In this present solution
Eq.(\ref{relaxationtime}) was used, which is consistent expression
for the relaxation time. Clearly, this is an indication that how the
physical scenarios implies in both solutions differ, depending on
the expression for the relaxation time. In summary, this solution
open the possibility to have effective phantom behaviors without
invoking exotic matter in a full causal thermodynamics formalism.

The phantom solution found presents a singularity which leads to
infinities in the energy density, pressure and also in the effective
pressure, temperature and entropy of the viscous fluid filling up
the universe in a finite time in the future. We have argue that the
obtained singularity requires to extend the previous classification
of singularities realized in~\cite{Nojiri2005}, in order to include
the behavior of the effective pressure which characterize non
perfect fluids.

An unexpected behavior of this solution is that the effective EoS is
constant and correspond to phantom matter, despite the dependence on
time of the thermodynamics parameters.

The accelerated expansion that present this solution implies that
the viscous stress is greater than the equilibrium pressure, so the
fluid has to be far from equilibrium. This is probably the main
criticism to the found solution. One can postulate that the causal
thermodynamics holds beyond the near-equilibrium regime, but there
are no consistent reasons to do this. Further investigations require
to face the consistency of the IS framework for accelerated
solutions and the restriction of near equilibrium, assumed in the
thermodynamical approaches. One first step in this direction is to
explore if a phantom solution of this type may exist in the
nonlinear generalization of causal thermodynamics developed
in~\cite{MaartensMendez}. We will explore this issue in a future
work.

\section*{Acknowledgments}

The authors acknowledge the comments and useful references given by
R. C. Nunes, Sergei Odintsov, Robert J. Scherrer and A. Vikman. N.
C. and S. L. acknowledge the hospitality of the Departamento de
Ciencias F\'{\i}sicas of Universidad de La Frontera where part of
this work was done. This work has been supported by Fondecyt grant
$N^{\circ}$ 1140238 (N. C.), VRIEA-DI-PUCV grant $N^{\circ}$
039.351/2016, Pontificia Universidad Cat\'{o}lica de Valpara\'{\i}so
(S. L.). The authors acknowledge M. Cruz for partial support in
Appendix.

\appendix*

\section{A}

\label{appendix}

If $s=1/2$, Eq.(\ref{eq:eq19}) adopts the form
\begin{widetext}
\begin{eqnarray}
\sqrt{3}\xi _{0}\left[ \frac{2}{3\left( 1-\omega ^{2}\right)
}\left( 3\left( 1+\omega \right)
\frac{\dot{H}}{H^{2}}+\frac{\ddot{H}}{H^{3}}\right) -3\right] +
\left[ 1+\frac{\sqrt{3}\xi _{0}}{2\left( 1-\omega ^{2}\right)
}\frac{1}{ \delta \left( \omega \right) }(\delta (\omega) -\frac{
\dot{H}}{H^{2}}) \right] \left[ 3\left( 1+\omega \right)
+2\frac{\dot{H }}{H^{2}}\right] =0, \label{eq:eqA1}
\end{eqnarray}
\end{widetext}
and this last equation admits a phantom solution. By using the
Ansatz given by Eq.(\ref{Ansatz}) in Eq.(\ref{eq:eqA1}), we obtain
the following algebraic equation for $A$
\begin{widetext}
\begin{eqnarray}
\sqrt{3}\xi _{0}\left[ \frac{2}{3A\left( 1-\omega ^{2}\right)
}\left( 3\left( 1+\omega \right) +\frac{2}{A}\right) -3\right] +
\left[ 1+\frac{\sqrt{3}\xi _{0}}{2\left( 1-\omega ^{2}\right)
}\frac{1}{ \delta \left( \omega \right) }\left( \delta \left(
\omega \right) -\frac{1}{A }\right) \right] \left( 3\left(
1+\omega \right) +\frac{2}{A}\right) =0, \label{eq:eqA3}
\end{eqnarray}
\end{widetext}
and if $0<\omega <1/2$ and $\sqrt{3}\xi _{0}>\left( 1-\omega
^{2}\right) /\left( 1/2-\omega \right) $, one of the solutions in
Eq.(\ref{eq:eqA3}) is real and positive.

\begin{figure}[h!]
\centering
\includegraphics[scale=0.19]{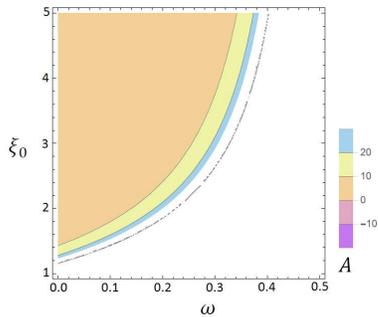}
\caption{Positive solution for $ A\left(\omega ,\xi _{0}\right)$
in the space parameters $\left(\omega ,\xi _{0}\right)$.}
\label{fig:1}
\end{figure}

In Fig.1 we show this positive solution as a function of $\omega $
and $\xi _{0}$.

\begin{figure}[h!]
\centering
\includegraphics[scale=0.19]{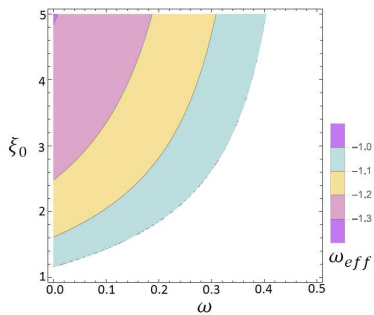}
\caption{The effective EoS parameter, $\omega _{eff}$, in the space
parameters $\left( \omega,\xi _{0}\right) $.} \label{fig:2}
\end{figure}
By other hand, according to Eq.(\ref{omegaefectivo1}), in Fig.2 we
show $\omega _{eff}\left( \omega ,\xi _{0}\right)
=-1-2/3A\left(\omega ,\xi _{0}\right)$ where $A\left( \omega ,\xi
_{0}\right)$ it is previous positive root.

We can see that it is possible to adjust $\omega_{eff}$ (phantom)
in the range of the observational data today ~\cite{Ade2014}
\begin{widetext}
\begin{equation}
\omega(0)=-1.019_{-0.080}^{+0.075}\;=\left\{
\begin{array}{c}
-0.944\;(\text{quintessence zone}),\\
-1.099\sim -1.1\;(\text{phantom zone}) .
\end{array}
\right.\label{eq:eqA4}
\end{equation}
\end{widetext}

\end{document}